\documentclass[paper,prd,reprint,preprintnumbers,nofootinbib,notitlepage,showpacs,showkeys]{revtex4-1}
\usepackage{latexsym,graphicx,amssymb,amsmath,mathrsfs} 
\usepackage{slashed}

\DeclareSymbolFont{bbold}{U}{bbold}{m}{n}
\DeclareSymbolFontAlphabet{\mathbbold}{bbold}

\newcommand{\be}{\begin{equation}}      
\newcommand{\ee}{\end{equation}}      
\newcommand{\bea}{\begin{eqnarray}}      
\newcommand{\eea}{\end{eqnarray}}    
     
\newcommand{\rt}[1]{{}}

\newcommand{\Tr}{\,\textrm{Tr}\,}

\newcommand{\ife}{\,\textrm{if}\,}
\newcommand{\nucl}{\,\textrm{nucl}\,}
\newcommand{\orr}{\,\textrm{or}\,} 
\newcommand{\ns}{\,\textrm{ns}\,}
\newcommand{\s}{\,\textrm{s}\,} 
\renewcommand{\min}{\,\textrm{min}\,} 
\newcommand{\GeV}{\,\textrm{GeV}\,} 
\newcommand{\MeV}{\,\textrm{MeV}\,} 
\newcommand{\els}{\,\textrm{else}\,}

\makeatletter
\renewcommand\appendix{\par
\setcounter{section}{0}%
\setcounter{subsection}{0}%
\gdef\thesection{\appendixname\space\@Alph\c@section}}

\long\def\unmarkedfootnote#1{{\long\def\@makefntext##1{##1}\footnotetext{#1}}}
\makeatother

\begin{document} 

\title{Mesonic and nucleon fluctuation effects at finite baryon density}
\author{G. Fej\H{o}s$^1$}
\email{fejos@rcnp.osaka-u.ac.jp}
\author{A. Hosaka$^{1,2}$}
\email{hosaka@rcnp.osaka-u.ac.jp}
\affiliation{$^1$Research Center for Nuclear Physics, Osaka University, Ibaraki, Osaka 567-0047, Japan}
\affiliation{$^2$Advanced Science Research Center, Japan Atomic Energy Agency, Tokai, Ibaraki, 319-1195 Japan}

\begin{abstract}
{Mesonic and nucleon fluctuation effects are investigated in medium. We couple the nucleon field to the $2+1$ flavor meson model and investigate the finite temperature and density behavior of the system, in particular, the axial anomaly function. Somewhat contrary to earlier expectations we find that it tends to strengthen at finite density. At lower temperatures nucleon density fluctuations can cause a relative difference in the $U_A(1)$ axial anomaly of about $20\%$. This has important consequences on the mesonic spectra, especially on the $\eta-\eta'$ system, as we observe no drop in the $\eta'$ mass as a function of the baryochemical potential, irrespective of the temperature. Based on the details of chiral symmetry restoration, it is argued that there has to be a competition between underlying QCD effects of the anomaly and fluctuations of the low energy hadronic degrees of freedom, and the fate of the $U_A(1)$ coefficient should be decided by taking into account both effects simultaneously.}
\end{abstract}

\pacs{11.30.Qc, 11.30.Rd}
\keywords{Axial anomaly, chiral symmetry breaking, functional renormalization group}  
\maketitle

\section{Introduction}

Chiral invariance, being an approximate symmetry of the theory of quantum chromodynamics (QCD), has played an important role in understanding the properties of light mesons both in vacuum and in medium. In particular, the $\eta'$ mass problem was solved via the discovery of the quantum anomaly of the $U_A(1)$ subgroup of chiral symmetry \cite{fukushima11}. Based on the underlying mechanism of topological fluctuations and instanton configurations of QCD, it is often argued that the axial symmetry should (partially) recover at finite temperature and/or baryon density \cite{schaefer98}. If so, it would affect the mesonic spectrum, and, in particular, the mass of the $\eta'$ meson could decrease by a significant amount. It is also argued that in nuclear medium, the chiral condensate drops about $30\%$ at the normal nuclear density \cite{jido08}, which could also imply a partial restoration of the $U_A(1)$ symmetry.

Earlier model calculations found that the reduction of the mass of the $\eta'$ meson can be of the order of $100 \MeV$ at normal nuclear density \cite{costa03,nagahiro06}, which effectively implies an attractive interaction between the $\eta'$ meson and the nucleon. Similarly to the $\Lambda(1405)$ resonance, which is strongly believed to be a $\bar{K}N$ bound state, this suggests that an $\eta'N$ bound state may also likely to be formed, given that the attraction between the two particles is indeed strong enough \cite{sakai13,sakai16}. Recently, there has been an increasing interest in searching for traces of such an object \cite{leps}.

Regarding the (partial) restoration of the axial anomaly at finite temperature and/or density, little is known about its nature. Lattice QCD results are controversial at the moment having different predictions in the restoration of the $U_A(1)$ subgroup at finite temperature \cite{dick15,sharma16,tomiya16}. Concerning effective model calculations for $2+1$ flavors, due to the melting of the chiral condensate at finite temperature, they predict a drop in the $\eta'$ mass around the critical temperature \cite{mitter14,heller16,rennecke16}, but a common feature of these calculations is to treat the anomaly parameter without being affected by fluctuations. In \cite{fejos16}, we introduced a scale-dependent anomaly function and investigated the effect of mesonic fluctuations (both quantum and thermal) on it. We found that these fluctuations are significant and cannot be neglected, as the axial anomaly tend to strengthen toward the critical temperature once they are taken into account.

Concerning the $\eta'-N$ interaction, despite experimental efforts \cite{moskal00,nanova12}, very little is known about its nature. In addition to the $U_A(1)$ anomaly, the interaction seems to be dictated by scalar meson exchange \cite{sakai13}, and one is also interested in the role of various vector mesons \cite{parganlija13,kovacs16}.

The main goal of this paper is to investigate the evolution of the $U_A(1)$ anomaly, and in addition to mesonic fluctuations, investigate the role of the nucleon at finite baryochemical potential. The most natural way to tackle the problem is to take the $2+1$ flavor linear sigma model ($L\sigma M$) and couple the meson field to the nucleon via a Yukawa type interaction \cite{sakai13,sakai16}. For the appropriate treatment of the short-range nucleon-nucleon interaction, the model can be extended with the $\omega$ and $\rho$ mesons \cite{drews13}, see also a recent review for the two-flavor version in \cite{drews16}. We note that even though these degrees of freedom are necessary to generate a first order nuclear liquid-gas transition, and thus important from the point of view of a proper description of nuclear matter, we do not include them in our framework. The reason is that our main focus is on the evolution of the `t Hooft coupling, and in a minimal nucleon-meson model, neither $\omega$ nor $\rho$ couples to the scalar (and pseudoscalar) mesons (and therefore to the anomaly) directly \cite{drews13,drews16}.
 
For the inclusion of quantum, thermal and density fluctuations, we use the functional renormalization group (FRG) method \cite{kopietz}. The used approximation scheme is the leading order of the derivative expansion, equipped with the so-called chiral invariant expansion technique \cite{fejos16}. The introduction of nucleons into the system allows us to investigate directly at finite nuclear density (or equivalently, at finite baryochemical potential $\mu_B$), and within the present method, we are able to obtain the fluctuation induced field, temperature, and chemical potential dependence of the anomaly coefficient.

The paper is organized as follows. In Sec. II, we introduce the model and the basics of the FRG method. In Sec. III, we review the approximation scheme and the details of the solution of our equations. The reader finds the main results and corresponding figures in Sec. IV, together with discussions on various aspects of our findings. Section V is dedicated for the conclusions.

\section{Model and method}

As announced in the Introduction, we are using the $2+1$ flavor linear sigma model coupled to a two-component isospinor describing the nucleon field. We neglect isospin asymmetry and thus, the neutron and the proton are treated on an equal footing. The dynamical fields that we are introducing is as follows. We denote the meson field by $M$,
\bea
M(x)=\sum_{a=0}^8 T^a\big(s^a(x)+i\pi^a(x)\big),
\eea
which belongs to a $U(3)$ Lie algebra [$T^a$ are $U(3)$ generators], and the $s^a$ and $\pi^a$ fields contain the scalar and pseudoscalar mesons, respectively. The nucleons are described by a two-component isospinor $\psi$, 
\bea
\psi^T(x)=\big(p(x),n(x)\big).
\eea
We introduce meson-nucleon couplings in the linear sigma model (L$\sigma$M) via a Yukawa-type interaction, and the action of the model is the following:
\bea
\label{Eq:lagr}
S&=&\int_x \Big(\Tr(\partial_\mu M^\dagger \partial^\mu M) - m^2 \Tr (M^\dagger M) \nonumber\\
&-& \frac{g_1}{9} \left(\Tr(M^\dagger M)\right)^2 - \frac{g_2}{3} \Tr (M^\dagger M M^\dagger M)\nonumber\\
&-&\Tr\big(H(M^\dagger+M)\big)-a (\det M^\dagger + \det M)\nonumber\\
&+&\bar{\psi}(i\slashed{\partial}-\mu_B\gamma_0-m_N)\psi-g\bar{\psi}\tilde{M}_5\psi\Big).
\eea
The first three lines correspond purely to the L$\sigma$M (note that no isospin breaking is present, thus $H=h_0T^0+h_8T^8$), which (apart from the explicit breaking term containing $H$) reflects $U_L(3)\times U_R(3)$ chiral symmetry via the transformation $M\rightarrow LMR^\dagger$, where $L$ and $R$ are independent $U(3)$ matrices. The fourth line contains the dynamics of the nucleon field, where $\mu_B$ denotes the baryochemical potential, and the coupling term to the mesons. We have six free parameters in the L$\sigma$M part (i.e., $m^2$, $g_1$, $g_2$, $a$, $h_0$, $h_8$), and two additional ones regarding the nucleons ($m_N$, $g$). Note that, the physical nucleon mass consists of two parts: the fermion mass term containing $m_N$, which breaks chiral symmetry explicitly, and a piece arising from the Yukawa interaction term, which leads to a nonzero contribution once spontaneous symmetry breaking occurred.

Since the mesons are described by a $3\times 3$ matrix, and $\psi$ is a two-component object, the last term in (\ref{Eq:lagr}) needs explanation. The nucleon has no strangeness; thus, we have to select an embedded $U(2)$ algebra in flavor $U(3)$ that corresponds to a purely isospin subalgebra ${\cal G_I}$. The way to do so is that we interchange the $T^0$ generator with the nonstrange one when defining ${\cal G_I}$. Note that the change from the 0$-$8 basis to ns$-$s (nonstrange$-$strange) is done via the ideal mixing
\bea
\label{Eq:TsTns}
\begin{pmatrix}
T^{\ns} \\ T^{\s}
\end{pmatrix}
= \frac{1}{\sqrt3} \begin{pmatrix}
\sqrt2 & 1 \\
1 & -\sqrt2
\end{pmatrix}
\begin{pmatrix}
T^0 \\ T^8
\end{pmatrix}.
\eea
The four matrices that span ${\cal G_I}$ is then $\{T^{\ns}, T^1, T^2, T^3\}$. The corresponding $\tilde{M}$ meson field will become effectively a $2\times 2$ matrix,
\bea
\tilde{M}=\sum_{a=\ns,1,2,3} T^a(s^a+i\pi^a),
\eea
which can couple to $\psi$, but we need the Lagrangian to be a scalar; thus, define
\bea
\tilde{M}_5=\sum_{a=\ns,1,2,3} T^a(s^a+i\gamma_5\pi^a).
\eea
This allows the combination $\bar{\psi}\tilde{M}_5\psi$ to have the appropriate transformation properties.

In this paper, we are interested in fluctuation effects of the mesons and the nucleon. These effects will be calculated in the language of the quantum effective action $\Gamma$, which is related to the Legendre transform of the logarithm of the partition function $Z$. For the set of fields $\Phi=(s^a,\pi^a,p,n)$, we have
\bea
Z[J]&=&\int {\cal D}\Phi e^{i(S+\int J\cdot\Phi)}, \nonumber \\
\Gamma[\Phi]&=&-i\log Z[J]-\int J\cdot\Phi,
\eea
where $J$ represents source fields. The fluctuations are included with the help of the functional renormalization group (FRG) method. In this framework, one defines a scale dependent effective action $\Gamma_k$, which includes fluctuations only with momenta $q\gtrsim k$, where $k$ is the so-called scale parameter. This is achieved by adding a regulator term to the classical action $S$,
\bea
\label{Eq:Smod}
S \quad \longrightarrow \quad S + \int_x\int_y \Phi^\dagger(x) {\cal R}_k(x,y) \Phi(y),
\eea
which (in Fourier space) can be interpreted as a momentum dependent mass term. Note that ${\cal R}_k$ is a matrix in accordance with the set of fields of $\Phi$. By choosing ${\cal R}_k$ such that it suppresses low momentum ($q \lesssim k$) fluctuations, while leaving high momentum ones ($q \gtrsim k$) unaffected, we can readily construct $\Gamma_k$. There are several types of regulator functions available; in this paper, we choose the so-called 3D Litim regulator \cite{litim01}, which, for given modes in Fourier space is as follows:
\bea
\label{Eq:Rbos}
R^B_k(q,p)=(k^2-{\bf q}^2)\Theta(k^2-{\bf q}^2)\delta({\bf q}+{\bf p})
\eea
for bosons (i.e., $s^a, \pi^a$), and
\bea
\label{Eq:Rfer}
R^F_k(q,p)=i{\bf p}\!\!\!/\Big(\sqrt{\frac{k^2}{{\bf p^2}}}-1\Big)\Theta(k^2-{\bf q}^2)\delta({\bf q}+{\bf p})
\eea
for fermions (i.e., $n, p$) [note that ${\bf p}$ and ${\bf q}$ are three-momenta]. The nice feature of this construction is that the effective action $\Gamma_k$ generated by (\ref{Eq:Smod}), and defined as
\bea
\Gamma_k[\Phi]&=&-i\log Z_k[J]-\int J\Phi-\int\int \Phi^\dagger {\cal R}_k \Phi
\eea
[$Z_k$ receives $k$ dependence via (\ref{Eq:Smod})] obeys the following flow equation \cite{kopietz}:
\bea
\label{Eq:flow}
\partial_k \Gamma_k = \frac{1}{2} \int_p \int_q \Tr[(-2)^F(\Gamma_k^{(2)}+{\cal R}_k)^{-1}(q,p) \partial_k {\cal R}_k(p,q)], \nonumber\\
\eea
where $\Gamma_k^{(2)}$ is the second functional derivative matrix of $\Gamma_k$ in Fourier space, and the factor of $(-2)^F$ indicates that when evaluating the trace, if one encounters a fermionic variable in $\Gamma_k^{(2)}$, it has to be taken with a negative sign due to their Grassmannian nature and a multiplicative symmetry factor of 2. Note that, at $k=0$, all fluctuations are included and thus, $\Gamma_{k=0}=\Gamma$, while at the highest (UV) scale $\Lambda$, as no fluctuations are present, it has to be equal to the classical action, $\Gamma_{k=\Lambda}=S$. 

Since (\ref{Eq:flow}) is an exact relation, it can only be solved in approximation schemes. Practically, one chooses an ansatz for $\Gamma_k$ and integrate (\ref{Eq:flow}) from $k=\Lambda$ to $k=0$ with the boundary condition $\Gamma_{k=\Lambda}=S$. If the theory in question is renormalizable, and it is expected that the UV cutoff does not interfere with the low energy behavior, then $\Lambda$ might also be chosen (formally) to be infinity. We, however, deal with an effective theory of the strong interaction and even though the model defined by (\ref{Eq:lagr}) is renormalizable, it makes no sense to apply an infinite UV cutoff limit. The L$\sigma$M is expected to be valid up to $\sim {\cal O}(1 \GeV)$; thus, we employ $\Lambda=1\GeV$.

\section{Solution of the flow equation}

In this section, we discuss the approximate solution of the flow equation (\ref{Eq:flow}). First of all, we need an ansatz for $\Gamma_k$. We split it into three parts: $\Gamma_{k,M}^{M}$ and $\Gamma_{k,M}^N$ refers to mesonic interactions, which arise from mesonic- and nucleon fluctuations, respectively, while $\Gamma_{k,N}$ is purely the nucleon part of the effective action,
\bea
\label{Eq:gamma0}
\Gamma_k=\Gamma_{k,M}^M+\Gamma_{k,M}^N+\Gamma_{k,N}.
\eea
The reason why it is worth to separate $\Gamma_{k,M}^M$ and $\Gamma_{k,M}^N$ is that nucleon interactions with mesons with nonzero strangeness are omitted in our framework; thus, $\Gamma_{k,M}^M$ has to reflect the original $U_L(3)\times U_R(3)$ symmetry, but $\Gamma_{k,M}^N$ should only have $U_L(2)\times U_R(2)$ invariance.

We are going to work with the leading order of the derivative expansion. As for $\Gamma_{k,M}^M$, we also apply the chiral invariant expansion technique \cite{fejos16},
\bea
\label{Eq:gamma1}
\Gamma_{k,M}^M&=& \int_x \Big(\Tr(\partial_\mu M^\dagger \partial^\mu M)-V_{k,M}^M\Big),\nonumber\\
V_{k,M}^M&=&U_k(\rho_2)+C_k(\rho_2)\rho_4+\Tr[H_k(M^\dagger+ M)] \nonumber\\
&+&A_k(\rho_2) \rho_{\det}, 
\eea
where
\bea
\rho_2&=&\Tr[M^\dagger M], \nonumber\\
\rho_4&=&\Tr\big[M^\dagger M-\Tr(M^\dagger M)/3\big]^2, \nonumber\\
\rho_{\det}&=&\det M^\dagger+\det M
\eea
are invariants of $U_L(3)\times U_R(3)$ [note that the last one is not invariant under the $U_A(1)$ subgroup]. For $\Gamma_{k,M}^N$ and $\Gamma_{k,N}$, we simply assume a form compatible with the classical action but with $k$-dependent couplings,
\bea
\label{Eq:gamma2}
\!\!\Gamma_{k,M}^N=-\int_x V_{k,M}^M, \quad \!\!V_{k,M}^M&=&f_{1,k}\Tr[\tilde{M}^\dagger \tilde{M}]\nonumber\\
&+&f_{2,k}\Tr[\tilde{M}^\dagger\tilde{M}\tilde{M}^\dagger \tilde{M}],
\eea
\bea
\label{Eq:gamma3}
\!\!\!\!\!\!\!\!\!\!\!\!\!\!\!\!\!\!\!\! \Gamma_{k,N}=\int_x \Big(\bar{\psi}(i\slashed{\partial}&-&\mu_B\gamma_0)\psi-V_{k,N}\Big), \nonumber\\
V_{k,N}&=&m_{N,k}\bar{\psi}\psi+g_k\bar{\psi}\tilde{M}_5\psi. 
\eea
Collecting (\ref{Eq:gamma1}), (\ref{Eq:gamma2}), and (\ref{Eq:gamma3}), then substituting (\ref{Eq:gamma0}) to (\ref{Eq:flow}), by the use of the regulators (\ref{Eq:Rbos}) and (\ref{Eq:Rfer}), we arrive at the following equation for the complete effective potential $V_k\equiv V_{k,M}^M+V_{k,M}^N+V_{k,N}$:
\bea
\label{Eq:flow2}
\partial_k V_k &=& \frac{k^4}{6\pi^2}\Bigg[T\sum_{\omega_n} \sum_{a=0}^8 \sum_{l=s^a,\pi^a}\frac{1}{\omega_n^2+k^2+m^2_{l,k}} \nonumber\\
&-&4T\sum_{\omega_j} \sum_{b=p,n}\frac{1}{(\omega_j-i\mu_B)^2+k^2+m^2_{b,k}}\Bigg],
\eea
where $m^2_{s^a,k}$, $m^2_{\pi^a,k}$ refer to the eigenmasses of the scalar and pseudoscalar sectors (i.e., their physical masses at scale $k$), respectively, while $m^2_{p,k}$ and $m^2_{n,k}$ are those for the protons and neutrons (i.e., eigenvalues of the $m_{N,k}^2{\bf 1}+g_k^2\tilde{M}^\dagger \tilde{M}$ matrix). The momentum integrals have been evaluated at finite temperature $T$; thus, summations over $\omega_n=2\pi nT$ bosonic and $\omega_j=2\pi(j+1/2)T$ fermionic Matsubara frequencies also appeared. The masses, being second derivatives of $V_k$, depend on the scale parameter $k$ and also on the actual field value of $M$ (note that, we are interested only in homogeneous configurations). The right-hand side of (\ref{Eq:flow2}) has to be projected onto each operator in $V_{k,M}^M$, $V_{k,M}^N$, and $V_{k,N}$ to obtain individual flow equations for the coefficient functions $U_k$, $C_k$, $A_k$, and for the paramaters $H_k$, $f_{1,k}$, $f_{2,k}$, $m_{N,k}$ and $g_k$.

First, let us consider $V_{k,N}$. Note that, in (\ref{Eq:flow2}), we implicitly assumed that $\psi=0$. Had we considered a general background where both $M,\psi \neq 0$ (one should think of appropriate sources that imply them), the meson and nucleon propagators would have mixed due to the Yukawa coupling. Concerning our ansatz (\ref{Eq:gamma1}), this mixing would solely be responsible for the flow of $V_{k,N}$ (a diagrammatical evaluation is also possible \cite{rennecke16}), but here we do not  calculate it and thus, set $V_{k,N}\equiv V_{\Lambda,N}$. This also means that $m_{N,k}\equiv m_{N}$ and $g_{k}\equiv g$ (and also $m_{p/n,k}^2\equiv m_{p/n}^2$, accordingly). We note that the neglected terms are of the ${\cal O}(g^3)$ and within our parametrization, expected to be small.

We are definitely interested in the $k$ dependence of $V_{k,M}^M$ and $V_{k,M}^N$. Details of establishing the flow of $V_{k,M}^M$ can be found in \cite{fejos16}. Let us briefly recall the procedure. By definition, contributions in the r.h.s. of (\ref{Eq:flow2}) for $V_{k,M}^M$ come only from the first term in the bracket. First, one sets the anomaly to zero ($A_k=0$) and calculates the mass matrices $\partial^2V_{k,M}^M/\partial s^a\partial s^b$, $\partial^2V_{k,M}^M/\partial \pi^a\partial \pi^b$ in a background defined by $M=v_0T^0+v_8T^8$ [note that a similar basis change is possible as of (\ref{Eq:TsTns}) to obtain $v_{\ns/\s} \leftrightarrow v_{0/8}$]. At this point, it is important to emphasize that here we do not take into account contributions of $V_{k,M}^N$ when calculating the masses, as it corresponds to mesonic fluctuations induced solely by the fermions, and thus, they would ruin the three-flavor chiral symmetry of $V_{k,M}^M$. After diagonalization, one expands the mesonic part of the r.h.s. of (\ref{Eq:flow2}) around $v_8=0$ and identifies the flows of $U_k$ and $C_k$ (for the latter it is crucial to combine terms into the $\rho_4$ invariant), and finds that $H_k$ does not depend on $k$, $H_k\equiv H$. Then, as a second step, one turns on the anomaly and expands around the zero anomaly configuration. This consists of recalculating the mass matrices in the presence of $A_k$ and keeping in the first term of the r.h.s. of (\ref{Eq:flow2}), the leading order piece around $A_k=0$. The procedure leads to the formation of the $\rho_{\det}$ invariant, and thus, one obtains the flow of $A_k$. The reader is referred to Appendix A for detailed expressions of $\partial_k U_k$, $\partial_k C_k$, $\partial_k A_k$, and for field derivatives of the action. Regarding the procedure in more detail, the reader should consult with \cite{fejos16}.

Now we are only left with the flow of $V_{k,M}^N$, i.e. the flows of couplings that correspond to the two-flavor chiral invariants made up by the $2\times 2$ matrix $\tilde{M}$, i.e., that of $f_{1,k}$ and $f_{2,k}$. These, by definition, come entirely from the second term in the bracket of the r.h.s. of (\ref{Eq:flow2}). The neutron and proton masses are
\bea
\label{Eq:m2pn}
\!\!\!\!m_{p/n}^2=m^2_{N}+\frac{g^2}{4}&&(s_{\ns}^2+s_1^2+s_2^2+s_3^2\nonumber\\
&&\!\!\!\!\!\!\!+\pi_{\ns}^2+\pi_1^2+\pi_2^2+\pi_3^2)\pm \frac{g^2}{2}\sqrt{\Delta},
\eea
where
\bea
\Delta&=&\pi_{\ns}^2\sum_{i=1,2,3} \pi_i^2+s_{\ns}^2\sum_{i=1,2,3} s_i^2\nonumber\\
&+&\frac12 \sum_{i,j,k=1,2,3}\epsilon_{ijk}^2\pi^2_i(s^2_j+s^2_k) \nonumber\\
&+&2\pi_{\ns}s_{\ns}\sum_{i=1,2,3} \pi_i s_i-\sum_{i\neq j=1,2,3} \pi_i\pi_js_is_j.
\eea
Inserting $m^2_{p/n}$ into the corresponding term in the flow equation (\ref{Eq:flow2}), we expand the obtained expression in terms of $g^2$ and arrive at
\bea
\label{Eq:flowVkMN}
\partial_k V_{k,M}^N&=&\frac{2k^4g^2}{3\pi^2}T\sum_{\omega_j}\frac{1}{\big[(\omega_j-i\mu_B)^2+k^2+m_N^2\big]^2}\nonumber\\
&\times&\Tr[\tilde{M}^\dagger\tilde{M}] \nonumber\\
&-&\frac{2k^4g^4}{3\pi^2}T\sum_{\omega_j}\frac{1}{\big[(\omega_j-i\mu_B)^2+k^2+m_N^2\big]^3}\nonumber\\
&\times&\Tr[\tilde{M}^\dagger\tilde{M}\tilde{M}^\dagger\tilde{M}] +{\cal O}(g^6).
\eea

This shows that via the complicated expressions of $m^2_{p/n}$ the r.h.s. of the flow equation indeed led to the formation of $\Tr[\tilde{M}^\dagger \tilde{M}]$ and $\Tr[\tilde{M}^\dagger \tilde{M}\tilde{M}^\dagger \tilde{M}]$, as expected from the ansatz (\ref{Eq:gamma3}). The flows of $f_{1,k}$ and $f_{2,k}$ are found to be
\begin{subequations}
\label{Eq:flowfs}
\bea
\label{Eq:flowf1}
\partial_k f_{1,k}&=&\frac{2k^4g^2}{3\pi^2}T\sum_{\omega_j}\frac{1}{\big[(\omega_j-i\mu_B)^2+E_k^2\big]^2}, \\
\label{Eq:flowf2}
\partial_k f_{2,k}&=&-\frac{2k^4g^4}{3\pi^2}T\sum_{\omega_j}\frac{1}{\big[(\omega_j-i\mu_B)^2+E_k^2\big]^3},
\eea
\end{subequations}
where the Matsubara sums can be performed analytically, see details in Appendix B.

The boundary condition for the effective action at the UV scale $k=\Lambda$ is $\Gamma_{\Lambda}=S$, thus
\bea
U_{\Lambda}(\rho_2)&=&m^2\rho_2+\frac{g_1+g_2}{9}\rho_2^2, \nonumber\\
C_{\Lambda}(\rho_2)&=&\frac{g_2}{3}, \qquad A_\Lambda(\rho_2)=a, \nonumber\\
H_{\Lambda}&=&H, \qquad f_{1,\Lambda}=0, \qquad f_{2,\Lambda}=0, \nonumber\\
m_{N,\Lambda}&=&m_N, \qquad g_\Lambda=g.
\eea
Now we are ready to solve the coupled equations (\ref{Eq:flow_Uk}), (\ref{Eq:flow_Ck}), and (\ref{Eq:flow_Ak}) together with (\ref{Eq:flowf1}) and (\ref{Eq:flowf2}). In the numerics all calculations are carried out in GeV units. For $U_k(\rho_2)$, $C_k(\rho_2)$, and $A_k(\rho_2)$, we set up grids in the interval $[0$:$2]$ with a step size of $10^{-2}$ and solve (\ref{Eq:flow_Uk}), (\ref{Eq:flow_Ck}) and (\ref{Eq:flow_Ak}) at each point. All necessary field derivatives are calculated with the seven-point formula, except for those close to the boundaries, where the five- and three-point formulas are used.  Equations (\ref{Eq:flowf1}) and (\ref{Eq:flowf2}) can be treated separately. In $k$ space we integrate all equations from the UV cutoff $\Lambda$ to zero using the Runge-Kutta algorithm.

\section{Results}

Once the effective action (\ref{Eq:gamma0}) is obtained, one is able to extract information on the mesonic spectrum (take the second derivatives), the vacuum expectation value for $v_{\ns}$ and $v_{\s}$ (search for the lowest energy configuration), the anomaly function $A_k$, etc. Note that in order to do so, it is required to calculate various field derivatives of the effective action. The necessary formulas can be found in Appendix B and in \cite{fejos16}. In this section, we review the numerical results that have been obtained at finite temperature and finite baryochemical potential.

\begin{figure}
\includegraphics[bb = 310 80 520 570,scale=0.33,angle=0]{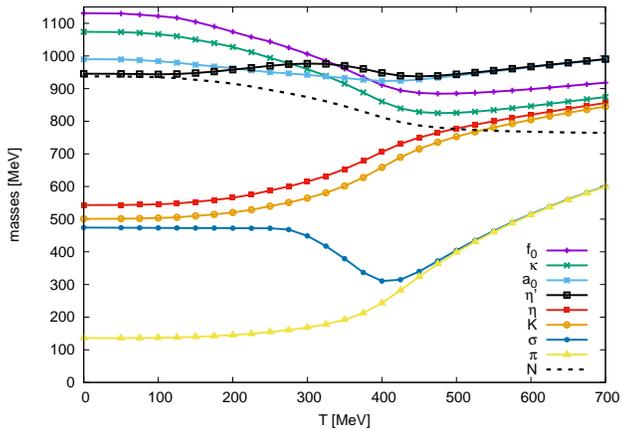}
\caption{Mass spectrum at finite temperature with $\mu_B=0$. The plot shows a similar behavior of the spectrum as of found in \cite{fejos16}. The $\eta'$ mass does not drop around the critical temperature.}
\label{fig1}
\end{figure}

The model needs to be parametrized, i.e., one has to determine the UV parameters in the vacuum (i.e., at $T=0$, $\mu_B=0$) via some physical input. First, we determine $h_0$ and $h_8$ from the partially conserved axialvector current (PCAC) relations (we use values of $f_\pi$, $f_K$ decay constants \cite{pdg}), in accordance with \cite{fejos16}, then fit the masses of $\pi$, $K$, $\eta$, $\eta'$, $N$, and finally, choose $g$ in a way that the nucleon mass gets as much contribution from the symmetry breaking as possible. The following values were employed for parametrization \cite{pdg}:
\bea
f_{\pi}&=&93 \MeV, \quad f_K=113 \MeV, \nonumber\\
M_{\pi}&=&140 \MeV, \quad M_K=494 \MeV, \nonumber\\
M_{\eta'}&=&958 \MeV, \quad M_\eta=548 \MeV, \nonumber\\
M_{\nucl}&=&939 \MeV.
\eea

\begin{table}[t]
\centering
\vspace{0.2cm}
  \begin{tabular}{ c | c }
    Parameter & Value \\ \hline
    $h_0$ & $ (286 \MeV)^3$ \\ \hline
    $h_8$ & $-(311 \MeV)^3$ \\ \hline
    $\mu^2$ & $-0.95 \GeV^2$ \\ \hline
    $g_1$ & $70$  \\ \hline
    $g_2$ & $160$  \\ \hline
    $a$ & $-3.0 \GeV$  \\ \hline  
    $g$ & $3.8$ \\ \hline
    $m_N$ & $755 \MeV$ \\ \hline
  \end{tabular}
  \caption{Set of parameters at the UV scale determined in the vacuum, i.e., at $T=0$, $\mu_B=0$.}
\end{table}

\begin{table}[t]
\centering
\vspace{0.2cm}
  \begin{tabular}{ c | c | c | c}
    Mass par. & $|A|_{\mu_B=0 \GeV}$ & $|A|_{\mu_B=0.5 \GeV}$ & $|A|_{\mu_B=1.0 \GeV}$ \\ \hline
    $m_N$ & 4.64\GeV & $4.64 \GeV$ & $5.19 \GeV$ \\ \hline
    $m_N/2$ & 4.38\GeV & $4.46\GeV$ & $6.09 \GeV$ \\ \hline
    $m_N/5$ & 4.28\GeV & $4.60\GeV$ & $6.19 \GeV$ \\ \hline
    $m_N/10$ & 4.29\GeV & $4.63\GeV$ & $6.12 \GeV$  \\ \hline
  \end{tabular}
  \caption{Absolute value of the `t Hooft coupling $|A|\equiv |A|_{k=0}$ in the minimum of the effective potential at $T=0$ for $\mu_B=0,0.5,1 \GeV$, respectively, varying the nucleon mass parameter.}
\end{table}

\begin{figure*}
\begin{center}
\raisebox{0.05cm}{
\includegraphics[bb = 280 80 520 570,scale=0.33,angle=0]{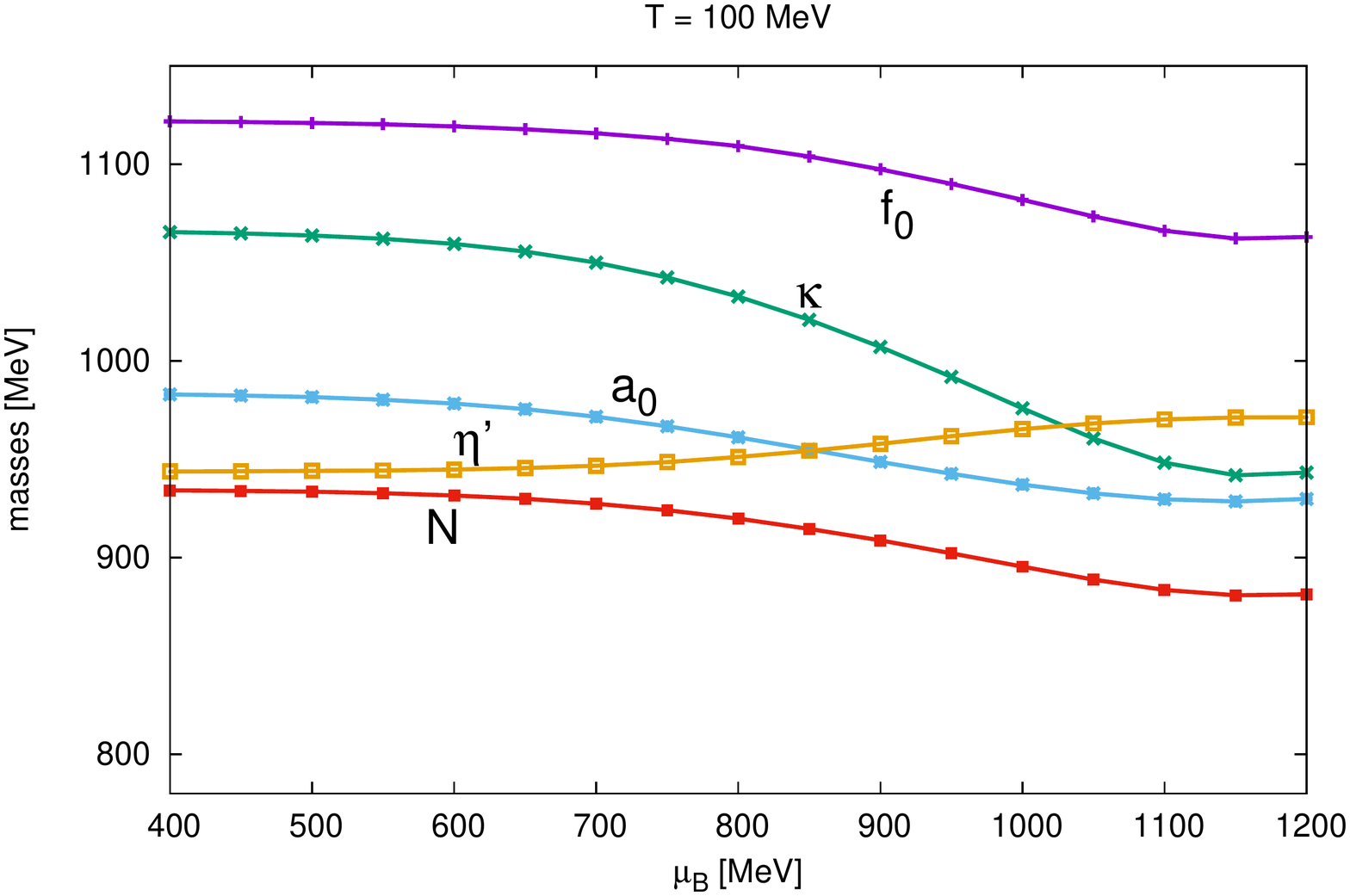}}
\includegraphics[bb = 1310 80 520 570,scale=0.33,angle=0]{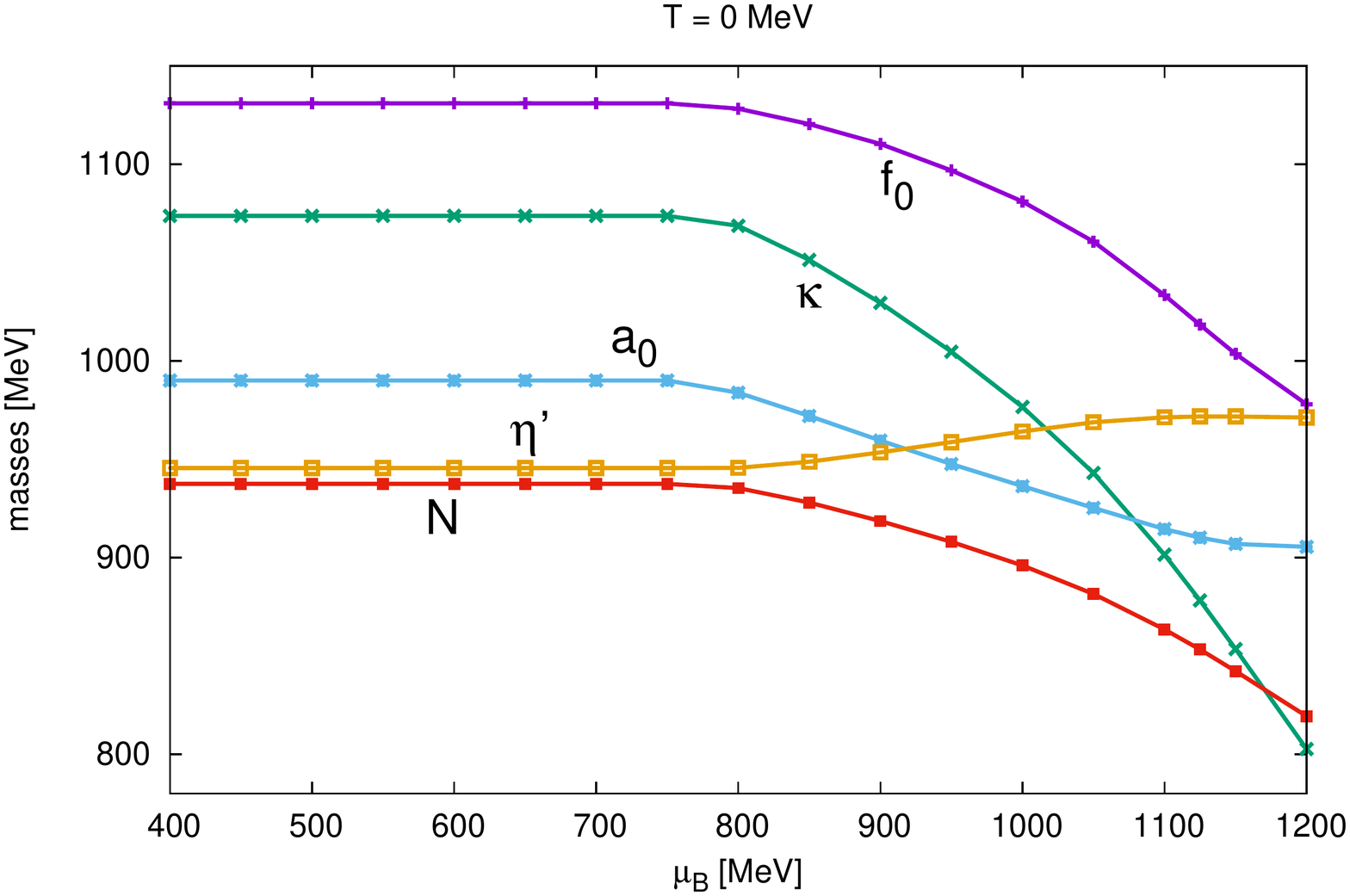}
\caption{Heavy meson spectrum at $T=0$ (left) and at $T=100\MeV$ (right) as a function of the chemical potential $\mu_B$.}
\label{fig2}
\end{center}
\end{figure*}

\begin{figure}
\includegraphics[bb = 310 80 520 570,scale=0.33,angle=0]{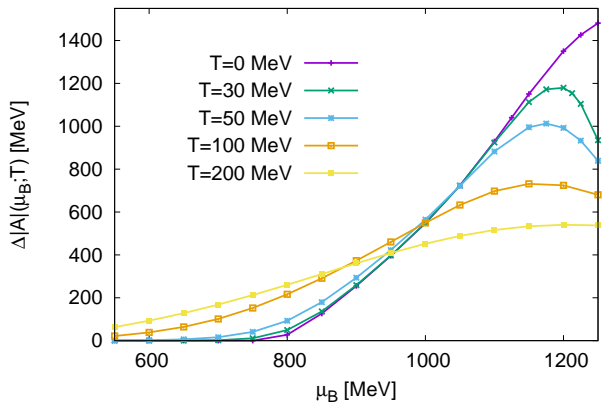}
\caption{Change of the anomaly in the minimum point of the effective potential as a function of $\mu_B$ at different temperatures.}
\label{fig3}
\end{figure}

\begin{figure*}
\begin{center}
\raisebox{0.05cm}{
\includegraphics[bb = 280 80 520 570,scale=0.33,angle=0]{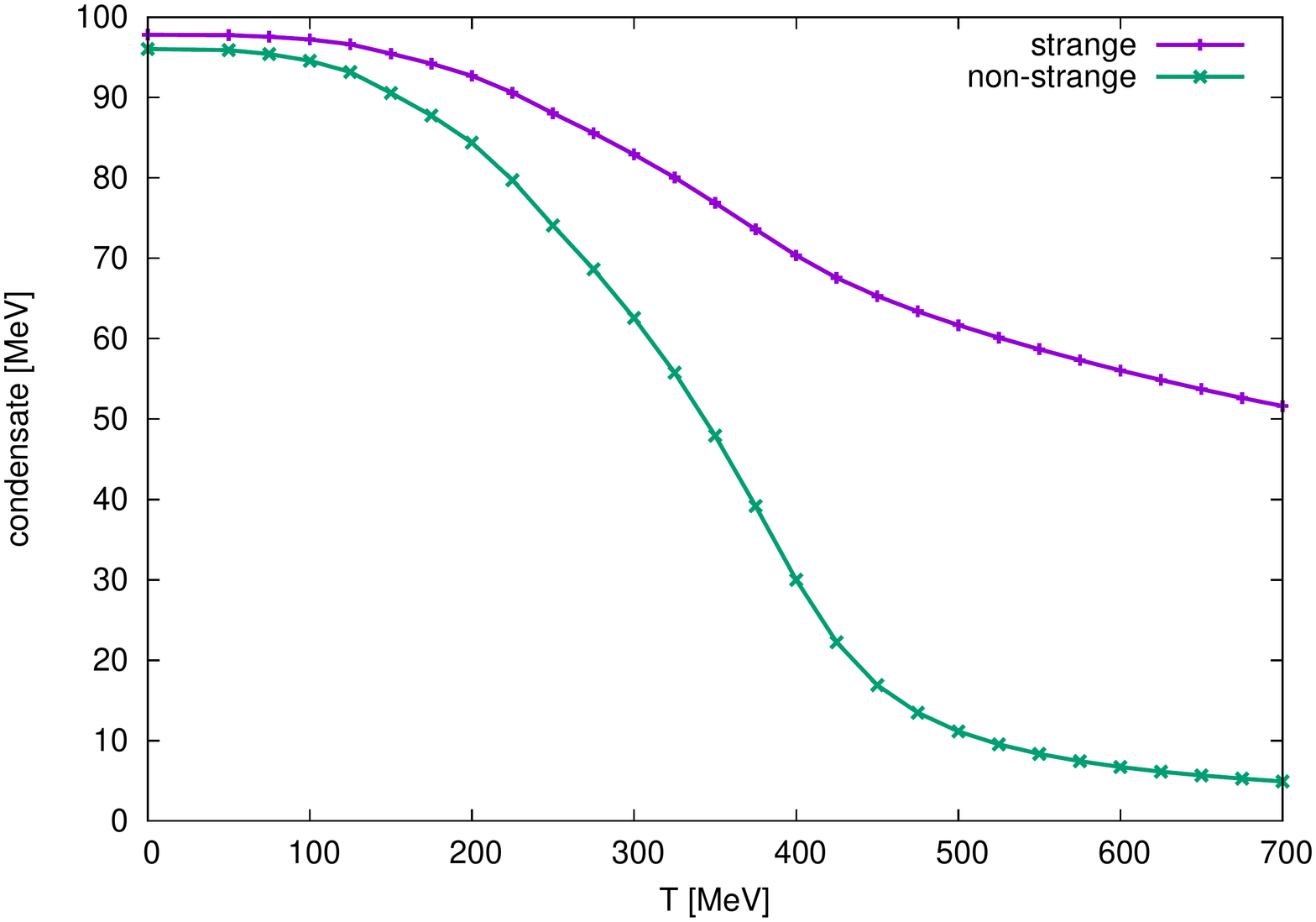}}
\includegraphics[bb = 1310 80 520 570,scale=0.33,angle=0]{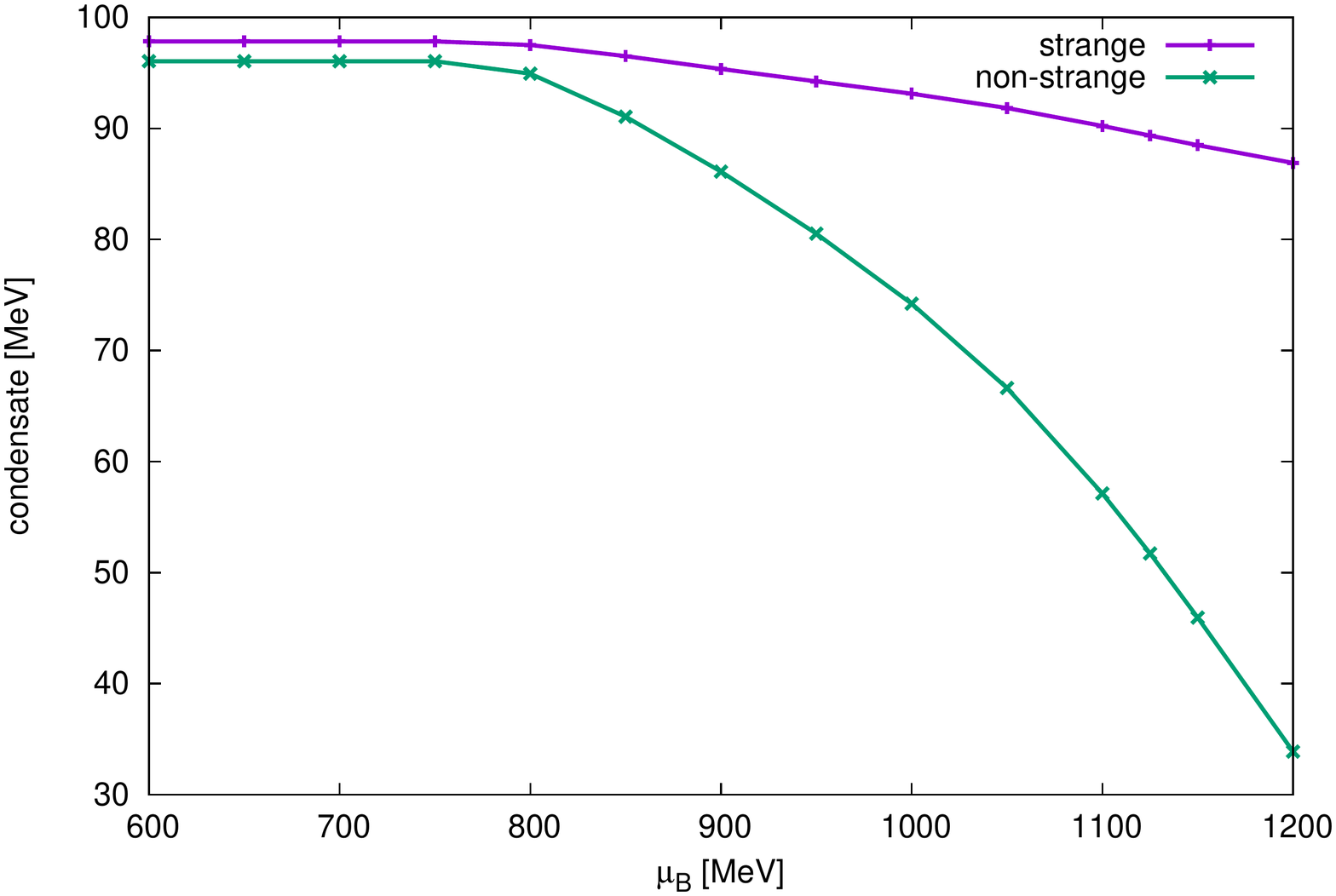}
\caption{(Partial) symmetry restoration at $\mu_B\neq 0$, $T=0$ (left) and at $\mu_B=0$, $T\neq 0$ (right). At $\mu_B=0$ the (pseudo)critical temperature $T_C$ is missed by a factor of $2$.}
\label{fig4}
\end{center}
\end{figure*}
One observes that the bare nucleon mass parameter, $m_N$, turns out to be quite large, see Table I for the details. The reason for the largeness of $m_N$ is that the mass component [see (\ref{Eq:m2pn})] that comes entirely from spontaneous symmetry breaking (i.e., $g^2v^2_{\ns}/4$) has a maximum as a function of $g$, and it is not enough to cover the most part of the nucleon mass. If $g$ is small, then obviously $gv_{\ns}$ goes to zero; however, if $g$ is large, nucleon fluctuations significantly backreact on the mesonic vacuum and push $v_{\ns}$ to a lower value. It is important to note that a large $m_N$ violates chiral symmetry explicitly, which is not expected based on the underlying theory of QCD. Because of the reason described above, we think of this violation as an artifact of the evolution of the $U_A(1)$ anomaly and the absence of instanton effects in the present approximation.
The reason why we do not think of it as a severe issue is that even though the largeness of $m_N$ certainly affects the fermionic fluctuation contributions quantitatively, we have done several runs by varying $m_N$, and obtained the same behaviors qualitatively. These are summarized in Table II.

In Fig. 1, we show the mass spectrum of the complete system (i.e., all the mesons and the nucleon itself) at finite temperature with zero chemical potential. The results are in accordance with the findings of \cite{fejos16}, the nucleon fluctuations do not produce any significant changes at zero density. We observe the peculiar behavior of the $\eta'$ mass, which shows no drop around the critical temperature due to the strengthening anomaly, as reported in \cite{fejos16}. 

One of the goals of the current study is to investigate the behavior of the spectrum at finite chemical potential. As mentioned in the Introduction, if the $\eta'$ mass had a drop of the order of $\sim 100 \MeV$ at the saturation density, one might expect the formation of an $\eta'N$ bound state due to an effective attractive interaction between the two particles. In Fig. 2, we show the heavier part of the mesonic spectrum at $T=0$ (left) and at $T=100 \MeV$ (right). First of all, we observe that a violation of the baryon Silver Blaze property \cite{cohen03} seems to occur, as at $T=0$, nucleon fluctuations start to show up not at the critical chemical potential, but at $\mu_B=m_N$, where $m_N$ is the nucleon mass parameter of the Lagrangian. This can be traced back to the expansion (\ref{Eq:flowVkMN}), where the structure of the summations clearly shows that at $T=0$, nonzero contributions do arise, if $\mu_B>m_N$. The violation, however, is not that severe, as up to the critical chemical potential $\mu_c = M_{\nucl} - B \approx 923 \MeV$ (here, $B$ denotes the binding energy), the change in the chiral condensates and masses are typically less than $\sim 10$\%. In terms of the FRG formalism, a possible resolution of the Silver Blaze violation was addressed in \cite{fu16}.

More importantly, on the contrary to usual expectations, the $\eta'$ mass does not decrease as the chemical potential is increased. This tendency does not depend qualitatively on the temperature, and it indicates that fluctuation effects push the $\eta'$ mass into a higher value, which again can be traced back to the high increase of the fluctuation corrected axial anomaly function [i.e., $A_{k=0}$]. This is shown in Fig. 3. We define an anomaly difference function $\Delta |A|$ as
\bea
\Delta |A|(\mu_B,T)=|A|_{k=0}(\mu_B,T)_{\min}-|A|_{k=0}(0,T)_{\min}, \nonumber\\
\eea
which gives account at a given temperature $T$ how the absolute value of the anomaly (in the minimum of the effective potential) changes due to nucleon density fluctuations, compared to its value at $\mu_B=0$. In the temperature range $T\simeq$ 0$-$200 MeV, the anomaly function $A_{k=0}$ in the minimum of the effective potential ($V_{k=0}$) is around $5 \GeV$; thus, Fig. 3 shows that density fluctuations can cause even up to a $20\%$ difference.

In Fig. 4 we show the behavior of the condensates at finite density and at finite temperature. We see again that at $T=0$, fluctuations show up at $\mu_B = m_N$, but by reaching $\mu_c$, the relative changes are less than $10\%$; thus, the SB violation remains small. Note that no first order chiral transition is observed beyond $\mu_c$, similarly as found in \cite{drews13} (the other jump in the order parameters, related to the nuclear liquid-gas transition, is in turn missing due to the absence of the dynamics of a neutral vector meson \cite{floerchinger12,weyrich15}).

As for the finite temperature case, it turns out that nucleon fluctuations further push $T_C$ to a higher value compared to the case of their absence \cite{fejos16}. This results in a $T_C$ that is missed by a factor of $2$ compared to recent lattice simulations \cite{borsanyi16}. The reason can be again found in the increase of the anomaly function, which leads us to the conclusion that the $U_A(1)$ effects of QCD, which are inexplicable in terms of the effective theory framework, are also important from the point of view of the finite temperature dynamics. These effects could be put in by hand via the temperature (and/or density) dependence of the bare anomaly coefficient $a$, and there should be a competition between mesonic and nucleon fluctuations versus the underlying $U_A(1)$ dynamics. The investigation of this issue is beyond the scope of the current study; nevertheless, what we have found is that both of them are necessary to be taken into account in order to reproduce experimental and lattice data properly.

\section{Conclusions}  

In this paper, we have investigated fluctuation effects in nuclear medium. Our main finding is that nucleon fluctuations are non-negligible at finite baryochemical potential $\mu_B$ and have a strong effect on the axial anomaly of the underlying theory of QCD. As a consequence, it turns out that the mass of the $\eta'$ meson does not decrease as a function of the chemical potential, somewhat contradictory to earlier assumptions. This phenomenon can be traced back to the fact that the axial anomaly function, which becomes field, temperature, and chemical potential dependent via mesonic and nucleon fluctuation effects, may increase up to 20\% as a function of $\mu_B$. 

The increasing mass of $\eta'$ may raise doubts on a possible $\eta'N$ bound state in nuclear medium, but a final conclusion should not be drawn at the moment. The reason is that even though we have demonstrated that nucleon fluctuations are non-negligible and produce an increasing anomaly at finite temperature $T$ and chemical potential $\mu_B$, the underlying instanton effects of QCD might also have a relevant contribution. This stems from the fact that 1) a large bare nucleon mass $m_N$ had to be introduced, because due to the strong anomaly, solely chiral symmetry breaking was not able to explain the physical nucleon mass and 2) the critical temperature $T_C$ is missed by a factor of $2$ at zero chemical potential, compared to recent lattice simulations. These problems could be circumvented by introducing a bare anomaly parameter at the level of the classical action that depends explicitly on both $T$ and $\mu_B$, representing underlying instanton dynamics.

Another important contribution may arise from explicit quark degrees of freedom, which was completely neglected in the current framework. Nevertheless, if we consider the theory valid up to ${\cal O}(1 \GeV)$, quarks may also play a significant role in the dynamics. Vector mesons, in particular, $\rho$ and $\omega$-meson exchange should also be included for a more complete treatment of the system. These extensions represent future works that will be reported elsewhere.

\section*{Acknowledgements}

G. F. would like to thank Antal Jakov\'ac for his remark on the zero temperature fermionic flows. This work was supported in part by Grants-in-Aid for Scientific Research from JSPS [Grant No. JP15H03663(B)].

\makeatletter
\@addtoreset{equation}{section}
\makeatother 

\renewcommand{\theequation}{A\arabic{equation}} 

\vspace{0.5cm}

\appendix 
\section{Flows in the mesonic sector}  

The purely mesonic component of the effective potential, $V_{k,M}^M$, is approximated as,
\bea
V_{k,M}^M&=&U_k(\rho_2)+C_k(\rho_2)\rho_4+\Tr[H_k(M^\dagger+M)]\nonumber\\
&+&A_k(\rho_2)\rho_{\det},
\eea
see (\ref{Eq:gamma1}). Following the procedure of \cite{fejos16}, and projecting the mesonic part of the flow equation (\ref{Eq:flow2}) onto each operator, one derives evolution equations for $U_k(\rho_2)$, $C_k(\rho_2)$, $A_k(\rho_2)$, and $H_k$,
\begin{widetext}
\bea
\label{Eq:flow_Uk}
\partial_kU_k(\rho_2)&=&\frac{k^4T}{6\pi^2}\sum_{n=-\infty}^\infty\Bigg[\frac{9}{\omega_n^2+E_\pi^2}+\frac{8}{\omega_n^2+E_{a_0}^2}+\frac{1}{\omega_n^2+E_\sigma^2}\Bigg], \nonumber\\
\eea
\bea
\label{Eq:flow_Ck}
\partial_k C_k(\rho_2)&=&\frac{k^4T}{6\pi^2}\sum_{n=-\infty}^\infty\Bigg[\frac{4(3C_k+2\rho_2C_k')^2/3}{(\omega_n^2+E_{a_0}^2)^2(\omega_n^2+E_\sigma^2)}
+\frac{128C_k^5\rho_2^3/9}{(\omega_n^2+E_\pi^2)^3(\omega_n^2+E_{a_0}^2)^3}+\frac{24C_k\left(C_k-\rho_2C_k'\right)}{(\omega_n^2+E_{a_0}^2)^3}\nonumber\\
&+&\frac{4\left(3C_kC_k'\rho_2+4\rho_2^2C_k'^2+C_k(3C_k-2C_k''\rho_2^2)\right)/3}{(\omega_n^2+E_{a_0}^2)(\omega_n^2+E_\sigma^2)^2}
+\frac{64C_k^3\rho_2^2(C_k-\rho_2C_k')/3}{(\omega_n^2+E_\pi^2)^2(\omega_n^2+E_{a_0}^2)^3}-\frac{48C_k^2\rho_2^2C_k'}{(\omega_n^2+E_\pi^2)(\omega_n^2+E_{a_0}^2)^3} \nonumber\\
&+&\frac{6C_k-17\rho_2C_k'}{(\omega_n^2+E_{a_0}^2)^2}\frac{1}{\rho_2}-\frac{6C_k+9\rho_2C_k'+2\rho_2^2C_k''}{(\omega_n^2+E_\sigma^2)^2}\frac{1}{\rho_2}
+\frac{4C_k(6C_k+9\rho_2C_k'+2\rho_2^2C_k'')/3}{(\omega_n^2+E_{a_0}^2)(\omega_n^2+E_\sigma^2)^2}\Bigg],
\eea
\bea
\label{Eq:flow_Ak}
\partial_k A_k(\rho_2)&=&\frac{k^4T}{6\pi^2}\sum_{n=-\infty}^\infty\Bigg[-\frac{9A_k'}{(\omega_n^2+E_\pi^2)^2}-\frac{9A_k}{\rho_2(\omega_n^2+E_\pi^2)^2}
-\frac{8A_k'}{(\omega_n^2+E_{a_0}^2)^2}+\frac{12A_k}{\rho_2(\omega_n^2+E_{a_0}^2)^2} \nonumber\\
&-&\frac{3A_k}{(\omega_n^2+E_\sigma^2)^2\rho_2}+\frac{7A_k'}{(\omega_n^2+E_\sigma^2)^2}+\frac{2\rho_2A_k''}{(\omega_n^2+E_\sigma^2)^2}\Bigg],
\eea
and $\partial_k H_k = 0$, where
\bea
E_\pi^2&=&k^2+U_k'(\rho_2), \nonumber\\
E_{a_0}^2&=&k^2+U_k'(\rho_2)+\frac43 \rho_2 C_k(\rho_2), \nonumber\\
E_{\sigma}^2&=&k^2+U_k'(\rho_2)+2\rho_2 U_k''(\rho_2),
\eea
and $\omega_n=2\pi nT$ denote bosonic Matsubara frequencies. The summations can be done analytically, see details in \cite{fejos16}. 

\renewcommand{\theequation}{B\arabic{equation}} 
\section{Fermionic effects}  

In the effective action $\Gamma_k$, nucleon induced mesonic interactions are represented by $\Gamma_{k,M}^N$. The corresponding flow equations (\ref{Eq:flowfs}) of the couplings $f_{1,k}$ and $f_{2,k}$ contain fermionic Matsubara sums. They read as
\bea
T\sum_j\frac{1}{[(\omega_j-i\mu_B)^2+E_k^2]^2}&=&\!\sum_\pm\!\Bigg[\frac{\tanh\Big(\frac{E_k\pm\mu_B}{2T}\Big)}{8E_k^3}-\frac{\cosh^{-2}\Big(\frac{E_k\pm\mu_B}{2T}\Big)}{16E_k^2T}\Bigg], \\
T\sum_j\frac{1}{[(\omega_j-i\mu_B)^2+E_k^2]^3}&=&\!\sum_\pm\!\Bigg[\frac{3\tanh\Big(\frac{E_k\pm\mu_B}{2T}\Big)}{32E_k^5}-\frac{3\cosh^{-2}\Big(\frac{E_k\pm\mu_B}{2T}\Big)}{64E_k^4T}-\frac{\cosh^{-2}\Big(\frac{E_k\pm\mu_B}{2T}\Big)\tanh\Big(\frac{E_k\pm\mu_B}{2T}\Big)}{64E_k^3T^2}\Bigg].
\eea
At $T=0$, appropriate limits are
\bea
T\sum_j\frac{1}{[(\omega_j-i\mu_B)^2+E_k^2]^2}\bigg|_{T=0}&=&\frac{\Theta(E_k-\mu_B)}{4E_k^3}-\frac{\delta(E_k-\mu_B)}{4E_kk}, \\
T\sum_j\frac{1}{[(\omega_j-i\mu_B)^2+E_k^2]^3}\bigg|_{T=0}&=&\frac{3\Theta(E_k-\mu_B)}{16E_k^5}-\frac{3\delta(E_k-\mu_B)}{16E_k^4}+\frac{\delta'(E_k-\mu_B)}{16E_k^3}.
\eea
\end{widetext}
Furthermore, one also needs field derivatives of the effective potential in order to calculate the nucleon fluctuation corrected meson masses and the condensate values of minimum energy. The corresponding formulas for $V_{k,M}^M$ can be found in \cite{fejos16}, here, we take care of $V_{k,M}^N$. One obtains the following expressions:
\renewcommand{\theequation}{B\arabic{equation}} 
\begin{widetext}
\bea
\frac{\partial V_{k,M}^N}{\partial s_i}\bigg|_{v_0,v_8}&=&
\begin{cases}
\frac{\sqrt2 f_{1,k}}{3}(\sqrt2 v_0+v_8)+\frac{f_{2,k}}{9\sqrt2}(\sqrt2 v_0+v_8)^3, \hspace{0.8cm} \ife \hspace{0.1cm} i=0\\
\frac{f_{1,k}}{3}(\sqrt2v_0+v_8)+\frac{f_{2,k}}{6}(\sqrt2 v_0+v_8)^3, \hspace{1.2cm} \ife \hspace{0.1cm} i=8\\
0, \hspace{6.4cm}  \els
\end{cases}
\eea
for the relevant first derivatives, and
\begin{subequations}
\bea
\frac{\partial^2 V_{k,M}^N}{\partial s_i s_j}\bigg|_{v_0,v_8}&=&
\begin{cases}
\frac13 \big(2f_{1,k}+f_{2,k}(2v_0^2+2\sqrt2 v_0v_8+v_8^2)\big),  \hspace{1.25cm} \ife \hspace{0.1cm} i=j=0\\
\frac{1}{3\sqrt2}\big(f_{1,k}+f_{2,k}(2v_0^2+2\sqrt2 v_0v_8+v_8^2)\big),  \hspace{1.05cm} \ife \hspace{0.1cm} i=0,\hspace{0.1cm} j=8 \hspace{0.1cm} \orr \hspace{0.1cm} i=8,\hspace{0.1cm} j=0\\
\frac16\big(2f_{1,k}+f_{2,k}(2v_0^2+2\sqrt2 v_0v_8+v_8^2)\big),  \hspace{1.25cm} \ife \hspace{0.1cm} i=j=8\\
f_{1,k}+f_{2,k}(v_0^2+\sqrt2 v_0v_8+v_8^2/2), \hspace{2.0cm} \ife \hspace{0.1cm} i=j=1,2,3\\
0, \hspace{6.65cm}  \els\\
\end{cases}\\
\frac{\partial^2 V_{k,M}^N}{\partial \pi^i \pi^j}\bigg|_{v_0,v_8}&=&
\begin{cases}
\frac{1}{18}\big(6(3+\sqrt2)f_{1,k}+f_{2,k}([6+4\sqrt2]v_0^2+[8+6\sqrt2]v_0v_8+[3+2\sqrt2]v_8^2)\big),  \hspace{0.3cm} \ife \hspace{0.1cm} i=j=0\\
\frac16\big(6f_{1,k}+f_{2,k}(2v_0^2+2\sqrt2 v_0v_8+v_8^2)\big), \hspace{5.3cm} \ife \hspace{0.1cm} i=j=1,2,3\\
0. \hspace{10.7cm}  \els
\end{cases}
\eea
\end{subequations}
for the second derivatives.
\end{widetext}

\end{document}